\documentclass[aps,prl,groupedaddress,twocolumn]{revtex4-1}
\usepackage{amsmath}
\usepackage{amsfonts}
\usepackage{physics}
\usepackage{graphicx}
\usepackage{mathtools}
\usepackage{amssymb}
\usepackage[colorlinks,linkcolor=blue,citecolor=blue,urlcolor=blue]{hyperref}
\DeclareMathOperator\arccosh{arccosh}
\DeclareMathOperator\sgn{sgn}
\DeclareMathOperator\Int{Int}
\begin{document}

\title{Dynamical Quantum Phase Transitions in Interacting Atomic Interferometers}
\author{Changyuan Lyu}
\author{Qi Zhou}
\email{zhou753@purdue.edu}
\affiliation{Department of Physics and Astronomy, Purdue University, 
West Lafayette, Indiana 47907, USA}
\date{\today}

\begin{abstract}
Particle-wave duality has allowed physicists to establish atomic interferometers as celebrated complements to their optical counterparts in a broad range of quantum devices. However, 
interactions naturally lead to decoherence and have been considered as a longstanding obstacle in implementing atomic interferometers in precision measurements. 
Here, we show that interactions lead to dynamical quantum phase transitions between Schr\"{o}dinger's cats in an atomic interferometer. These transition points result from zeros of Loschmidt echo, which approach the real axis of the complex time plane in the large particle number limit, and signify pair condensates, another type of exotic quantum states featured with prevailing two-body correlations. Our work suggests 
interacting atomic interferometers 
as a new tool for exploring dynamical quantum phase transitions and
creating highly entangled states to beat the standard quantum limit. 
\end{abstract}

\maketitle

Atomic interferometers have been playing crucial roles in modern quantum techniques. 
Their applications in precision measurements span a wide spectrum of problems, 
ranging from measuring the gravitational acceleration and the fine structure 
constant to detecting gravitational waves \cite{Fixler74, Fray2004, Pierre2006, Graham2013, Dimopoulos2008}.
The recent developments in ultracold atoms 
further prompt a precise control of atomic interferometers, including realizing highly tunable atomic beam splitters in a variety of systems \cite{Lawall1994, Glasgow1991, Pfau1993, Houde2000, Grimm94}
and accessing an atomic Hong-Ou-Mandel interferometer using 
optical tweezers \cite{HongOuMandel, RegalPRX, Regal2014Science, Lopes2015}. 
Despite the apparent particle-wave duality, there exists an intrinsic distinction
between atomic interferometers and their optical counterparts. Whereas many 
optical systems are essentially non-interacting, mutual interactions between particles
naturally exist and inevitably induce decoherence 
\cite{BuchleitnerDecoherence,ChalkerDecoherence,Jamison2011}, which poses
a grand challenge in implementing interferometers based on particles in precision
measurements.  

Dynamical quantum phase transition (DQPT) \cite{Heyl_PRL_2013, Heyl2017PRB, Zvyagin2016review, Heyl2018review} has recently invoked enthusiasm in 
multiple disciplines.  
It considers a particular type of Loschmidt echo, $|G(t)|^2$, where $G(t)=\langle \psi(0)|e^{-\frac{i}{\hbar}\hat{H}t}|\psi(0)\rangle$, 
$|\psi(0)\rangle$ is the initial state and $\hat H$ is the Hamiltonian controlling the time evolution of the quantum system. If one treats the time, $t$, as a tuning parameter, as analogous to the temperature or coupling strength in phase transitions at equilibrium, a vanishing $G(t)$ leads to a nonanalytic rate function 
$\lambda(t)\equiv-\lim_{N\rightarrow\infty}\frac{1}{N}\ln |G(t)|^2$, 
where $N$ is the number of degrees of freedom, and defines a critical time $t_c$. Fundamentally, DQPTs can be understood from zeros of $G(z)$ in the complex time plane by extending the real time, $t$, to the complex domain, $t\rightarrow z\equiv t+i\tau$. With increasing $N$, discrete zeros merge to continuous manifolds and eventually touch the real $t$ axis, making physical observables nonanalytic, similar to Lee-Yang zeros and Fisher zeros in the complex plane of the temperature or other parameters \cite{LeeYang, Fisher1978}. Whereas observations of DQPTs have been reported
in certain spin systems \cite{Heyl_PRL_2013, Heyl_PRL_2014, DQPT_exp_prl_2017, smale2018observation, Sharma2016, Bhattacharya2017},
such novel concept well deserves both theoretical and experimental studies in a much broader range of systems.  

In this Letter, we show that interactions in atomic interferometers could be turned into a unique means of creating highly entangled quantum states and exploring DQPTs between such states. 
Starting from a trivial initial state, where all bosonic atoms occupy the same quantum state in the interferometer, interactions give rise to intriguing quantum dynamics beyond the simple description of Rabi oscillations in non-interacting systems. Remarkably, DQPTs emerge as a result of zeros of $G(z)$ in the complex time plane approaching the real axis with the total particle number increased. Near a characteristic time scale that is inversely proportional to the interaction strength, there exist critical times, $t_c$, characterizing the transitions between different types of Schr\"{o}dinger's cats. Moreover, different from other DQPTs that have been studied in the
literature \cite{Heyl_PRL_2013,Heyl_PRL_2014, Heyl2015, DQPT_time_crystal, smale2018observation, Sharma2016, Bhattacharya2017, Karrasch2017}, 
$t_c$ here by itself corresponds to the rise of a pair condensate, a premier example of exotic condensate featured with vanishing one-body correlation function and prevailing two-body 
correlations \cite{James1982, Fischer2013pra, chen2017unfolding}. 
As for the dynamically generated Schr\"{o}dinger's cats, they are much more stable than those at equilibrium.  Since it is well known that Schr\"{o}dinger's cats allow physicists to beat the standard quantum limit in quantum measurements, our results suggest a new scheme of using non-equilibrium dynamics in interacting atomic interferometers to access highly entangled states for improving quantum sensing \cite{Giovannetti1330,Giovannetti2006,Giovannetti2011,Leibfried1476}. 
\begin{figure*}
    \includegraphics[width=\textwidth]{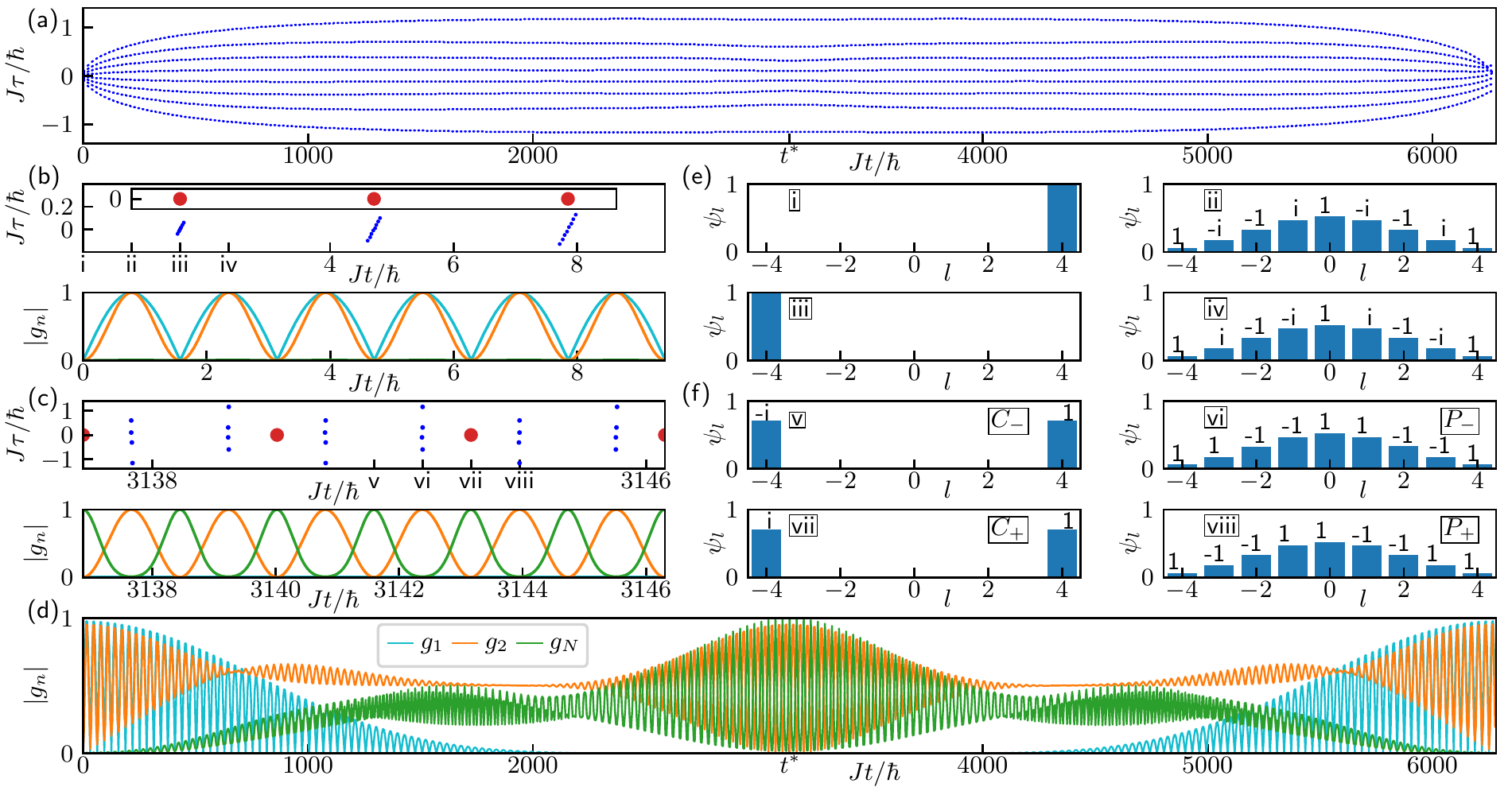}
    \caption{\label{fig:zeros_and_gn} (color online)
    Dynamics of $8$ bosons when $U/J=0.001$. 
    (a) Each blue dot represents a simple zero of $G(z)$ in the complex plane of time. 
    (d) Normalized correlation functions $\frac{2g_1}{N}, \frac{4g_2}{N(N-1)}, \frac{2g_N}{N!}$.
    (b) and (c) zoom into $0<t<3\pi$ and $|t-t^*|<3\pi/2$, respectively. Red dots in (b) and (c) are zeros of 
    $G(z)$ with multiplicity $8$ for the non-interacting case. (e) shows 
    the wave functions expanded by Fock states $\ket{\psi(t)} = \sum\psi_l\ket{\frac{N}{2}+l,\frac{N}{2}-l}$ at the four times, $Jt/\hbar = 0, \pi/4, \pi/2, 3\pi/4$.
    Numbers on top of bars are relative phases. (f) shows the wave functions at 
    $\tilde t_i $ (defined in Eq.\ \eqref{eq:states_and_times}). }
\end{figure*}

\paragraph{Hamiltonian.}
We consider $N$ bosonic atoms in an interferometer consisting of two quantum states. 
A generic Hamiltonian describing beam splitters in atomic interferometers reads
    $\hat{H} = -J(\hat{a}_1^\dagger \hat{a}_2 + \hat{a}_2^\dagger \hat{a}_1) 
+ g(\hat{n}_1^2+\hat{n}_2^2)+2g_{12}\hat{n}_1\hat{n}_2$,
where $J>0$ is the 
the coupling strength between the two quantum states, $\hat{a}_{i}^\dagger$ is the creation operator in the $i$th state, and $\hat{n}_i=\hat{a}^\dagger_i\hat{a}_i$.  $g$ and $g_{12}$ are the intra- and inter-state interactions, respectively. This Hamiltonian can be rewritten as 
\begin{equation}\label{eq:H}
   \hat{H} = -J(\hat{a}_1^\dagger \hat{a}_2 + \hat{a}_2^\dagger \hat{a}_1) 
    + \frac{\bar{U}}{2}(\hat{n}_1+\hat{n}_2)^2+\frac{U}{4}(\hat{n}_1-\hat{n}_2)^2, 
\end{equation} 
where $\bar{U}=g+g_{12}$, $U=2(g-g_{12})$. Due to the conservation of the total particle number $N=n_1+n_2$, $\bar{U}$ only contributes a trivial total phase of the wave function in the dynamics. We thus focus on interaction effects caused by $U$.  In the absence of $U$, Eq.\ \eqref{eq:H} corresponds to a beam splitter for non-interacting particles.
In the presence of interactions, though this Hamiltonian has been well studied 
\cite{Milburn1997, Ho2004, BuchleitnerDecoherence, Salgueiro2007, Salgueiro2007}, all our results, including zeros of $G(z)$ in the complex time plane, DQPTs, dynamically generated Schr\"{o}dinger's cats and pair condensates, elude the literature. Here, we solidify the discussion for repulsive interactions, $U>0$. Attractive interactions lead to similar results (Supplemental Materials).

\paragraph{Zeros in the complex plane.} We consider an initial state, $\ket{\psi(0)} = \ket{N,0}=\frac{1}{\sqrt{N!}}\hat a_1^{\dagger N}|0\rangle$, where all bosons occupy the same quantum state. The dynamical evolution, $\ket{\psi(t)}=e^{-\frac{i}{\hbar}\hat{H}t}\ket{\psi(0)}$,  is computed by expanding $\ket{\psi(0)}$ using the exact eigenstates of $\hat H$. Whereas this can be done for any parameters, we consider
$UN^{2}\ll J$. Such energy scale separation leads to a time scale separation,
\begin{equation}
    T\equiv\frac{\pi\hbar}{J} \ll t^*\equiv\frac{\pi\hbar}{U},
\end{equation}  
which allows us to access intriguing quantum dynamical evolutions
exhibiting extraordinary features.
When $U$ vanishes, the quantum dynamics is simply governed by 
\begin{subequations}
    \label{eq:operator_substitution}
    \begin{align}
       & \hat{a}_1^\dagger \rightarrow \cos\frac{Jt}{\hbar} \hat{a}_1^\dagger + i\sin\frac{Jt}{\hbar} \hat{a}_2^\dagger,\\
        &\hat{a}_2^\dagger \rightarrow i\sin\frac{Jt}{\hbar} \hat{a}_1^\dagger + \cos\frac{Jt}{\hbar} \hat{a}_2^\dagger.
    \end{align}
\end{subequations}
Thus, $
|\psi^{o}(t)\rangle=\frac{1}{\sqrt{N!}}(\cos(Jt/\hbar) \hat{a}_1^\dagger + i\sin(Jt/\hbar) \hat{a}_2^\dagger)^{ N}|0\rangle$, 
where the superscript $o$ represents the result of a non-interacting system. Extending $t$ to the complex plane,
it is straightforward to evaluate $G(z)$ and obtain its zeros. All zeros of $G(z)$ are located on the real axis. 
When $z=t^{o}_k\equiv(k+1/2)T$, where $k$ is an integer, the quantum many-body state becomes $\ket{0,N}=\frac{1}{\sqrt{N!}}\hat a_2^{\dagger N}|0\rangle$, and $G(t^{o}_k)=0$. 
This is expected, as in non-interacting systems, one can view each identical boson as a spin-$1/2$ rotating about an effective transverse magnetic field given by $J$. All spin-$1/2$s initially at the north pole of the Bloch sphere move to the south pole at the same times $t_k^{o}$, leading to a vanishing $G(z)$. 

Turning on a weak interaction that satisfies $UN^2\ll J$, one may expect that its effects are small. As shown in Fig.\ \ref{fig:zeros_and_gn}(b), this is indeed the case at small times. A given multiple zero with multiplicity $N$ now splits into $N$ simple zeros. Nevertheless, these zeros are close to each other and do not deviate much from the zeros of non-interacting systems, reflecting the perturbative role of a weak interaction at small times.  Indeed, the expansion of $|\psi(t)\rangle$ using Fock states is very similar to that of a non-interacting case, as shown in the four panels of Fig.\ \ref{fig:zeros_and_gn}(e).  For instance, at time $t=t^{o}\pm T/4$, $|\psi(t)\rangle$ is well represented by $\frac{1}{\sqrt{2^NN!}}(\hat a_1^\dagger\pm i \hat a_2^{\dagger })^N|0\rangle$, which corresponds to a binomial distribution when expanded by the Fock states $|l\rangle\equiv|N/2+l,N/2-l\rangle$. To simplify notations, we consider even $N$ here. See Supplemental Materials for results of odd $N$. However, at large times, even a weak interaction has profound effects. The separation between different zeros of $G(z)$ gets amplified greatly. In particular, near $t^*$, these zeros deviate largely from those of non-interacting systems. Whereas such zeros have finite imaginary parts, they intrinsically affect physical observables in the real time axis, as shown later.

\paragraph{Dynamically generated entangled states.} To further reveal the quantum
states emerged from this non-equilibrium dynamics and their intrinsic relations to 
the zeros of $G(z)$, we evaluate generic $s$-body 
correlation functions in the real time axis, 
$g_s=\bra{\psi(t)} {\hat a_1^{\dagger s}\hat a_2^s}\ket{\psi(t)}$, where $s>0$. At $t=0$, 
the initial Fock state has vanishing $g_s$ for any $s$. As time goes on, $g_s$ 
increases as a result of tunnelings between the two quantum states. When $U=0$, 
the dynamics is fully captured by Rabi oscillations. When $U\neq 0$, 
as shown in Fig.\ \ref{fig:zeros_and_gn}(d), one-body correlation function, $g_1(t)$, 
decays due to interaction induced decoherence. However, higher order correlation
functions have distinct behaviors. Normalized two-body and N-body correlation functions, 
$\frac{4g_2(t)}{N(N-1)}$ and $\frac{2g_N(t)}{N!}$,
reach their maxima around $t=t^*$. 
In the vicinity of $t^*$, both $|g_2|$ and $|g_N|$ oscillate  
with a period $T/2$.  This indicates the rise of highly entangled states with multi-particle correlations. 
Indeed, as shown in Fig.\ \ref{fig:zeros_and_gn}(c,f),  the following four states showing up alternatively near $t^*$ can be well captured by
\begin{align}\label{eq:states_and_times}
\begin{aligned}
    \tilde t_0 &=k T, & \ket{C_-}&=\frac{\hat{a}^{\dagger N}_1-i\hat{a}^{\dagger N}_2}{\sqrt{2N!}}\ket{0}, \\
    \tilde t_1 &=(k+\frac{1}{4})T, & \ket{P_-}&=\sum_{n=0}^N  \frac{i^{N-n}-i^{n+1}}{p_n}\hat{a}^{\dagger n}_1\hat{a}^{\dagger N-n}_2\ket{0}, \\
    \tilde t_2 &=(k+\frac{2}{4})T, &\ket{C_+}&=\frac{\hat{a}^{\dagger N}_1+i\hat{a}^{\dagger N}_2}{i^{1-N} \sqrt{2N!}}\ket{0}, \\
    \tilde t_3 &=(k+\frac{3}{4})T, & \ket{P_+}&=\sum_{n=0}^N\frac{i^{N-n}+i^{n+1}}{i^{1-N}\cdot p_n} \hat{a}^{\dagger n}_1\hat{a}^{\dagger N-n}_2\ket{0},
\end{aligned}
\end{align}
where $\tilde t = t-t^*$ and $p_n = n!(N-n)!\sqrt{\frac{2^{N+1}}{N!}}$. $\ket{C_\pm}$ are Schr\"{o}dinger's cats with vanishing $g_{s<N}$ and $|g_{N}|=N!/2$. We have verified that any $g_{s<N}$ does vanish when  Schr\"{o}dinger's cats arise. For clarity of the plots,  $g_{2<s<N}$ are not shown in the figure. $\ket{P_\pm}$ are pair condensates with $g_1=0$ and $|g_2|=N(N-1)/4$.  

The origin of emergent Schr\"{o}dinger's cats in the time domain can be traced back to the energy spectrum in the limit $UN^2\ll J$ (Supplemental Materials), which is written as
\begin{align}\label{eq:energy_first_order}
    &E_n =An +Bn^2,\ n=0,1,..,N,\\
    &B=-\frac{U}{2}, \quad A=\frac{UN}{2}+2J, \quad r\equiv\frac{A}{B}.
\end{align}
For any initial state $\ket{\psi(0)} = \sum_{n=0}^N c_n\ket{E_n}$, the wave function at a later time is given by $\ket{\psi(t)} = \sum_{n=0}^N c_n e^{-\frac{i}{\hbar}E_nt}\ket{E_n}$. Tuning $J$ and $U$,  when $r=r_m$ is satisfied, where $ r_m=4m+2$ or $4m$, $m\in\mathbb{Z}$, $\ket{C_\pm}$ can be easily identified. 
If $r=4m$, we obtain
\begin{equation}
\ket{\Psi(t^*)} = \sum_{n=0}^N c_n e^{-\frac{i}{\hbar}E_nt^*}\ket{E_n}=\sum_{n=0}^N c_n\frac{1-i(-1)^n}{\sqrt 2}\ket{E_n}.\label{ps}
\end{equation}
Since the energy eigenstates have well defined parity, 
\begin{align}\label{eq:property_of_eigenstates}
    \hat P\ket{E_n}=(-1)^n\ket{E_n},
\end{align} 
where $\hat P$ is the inversion operator, $\hat P\ket{l}=\ket{-l}$ and $[\hat{H}, \hat{P}]=0$. Using Eq.\ (\ref{ps}) and (\ref{eq:property_of_eigenstates}), we conclude that $\ket{\psi(t^*)}=(\ket{\psi(0)}-i\hat{P}\ket{\psi(0)})/\sqrt{2}$. Whereas this result is valid for any initial state, the initial state we chose gives rise to $\ket{C_-}$ emerging at $t=t^*$. Meanwhile, interaction effects are negligible in a short time scale of a few $T$s. The time evolution in such time scale is well captured by Eq.\ (\ref{eq:operator_substitution}) if we replace $t$ by $t-t^*$. Applying such transformation to $\ket{C_-}$,  it is straightforward to show that the other three states in Eq.\ \eqref{eq:states_and_times}  show up in corresponding times.  If  $r=4m+2$, the same discussions apply and the four states, $\ket{C_+}$,  $\ket{P_+}$, $\ket{C_-}$ and $\ket{P_-}$, show up at times 
$\tilde t_0, \tilde t_1, \tilde t_2, \tilde t_3$ in Eq.\ \eqref{eq:states_and_times}. It is also worth mentioning that, for odd particle numbers, the pair condensates are described by another type of wave functions $\sim \sum_l \psi_l'\hat a^{\dagger2l}_1 \hat a_2^{\dagger N-2l}|0\rangle$ (Supplemental Materials).

When $r\neq r_m$, Eq.\ (\ref{ps}) can not be satisfied. Nevertheless, the states near $t=t^*$ can be well approximated by Schr\"{o}dinger's cats in the weakly interacting regime. 
We  calculate the fidelity as a function of time,
\begin{align}
    \label{eq:cat_probability_definition}
    P(t) = \max(|\bra{{C}_+}\ket{\psi(t)}|^2, |\bra{{C}_-}\ket{\psi(t)}|^2).
\end{align}
Near $t^*$, we obtain
\begin{align}
\label{eq:Pt_summation_form}
\begin{aligned}
    &P(t)\approx \sqrt{\frac{1}{1+\frac{N^2}{4}(\frac{\pi}{2}-\frac{U}{2\hbar}t)^2}}\times\\
    &\sum_k 
    \left|\exp(-\frac{1}{\frac{2}{N}+i(\frac{\pi}{2}-\frac{U}{2\hbar}t)}(\frac{k\pi}{2}-\frac{\pi N}{4} - \frac{Jt}{\hbar})^2 )\right |^2.
\end{aligned}
\end{align}
Detailed calculations are presented in the Supplemental Materials. Near $t^*$,
$P(t)$ consists of multiple gaussian peaks centered at a series of discrete 
times with a separation $T/2$. 
Since the width of those peaks is about $\frac{\hbar}{\sqrt{N}J}$, only one peak contributes to $P(t)$ significantly at any $t$ in the large $N$ limit. $P(t)$ reaches its maximum at 
${t}^{*\prime}=\frac{k_0\pi\hbar}{2J}-\frac{\pi N\hbar}{4J}$, and 
\begin{align}
    \max[P(t)]=(1+(\frac{N\pi Ud}{8J})^2)^{-1/2},
\end{align}
where $k_0=\Int(\frac{2J}{U}+\frac{N}{2})$, the integer nearest to $\frac{2J}{U}+\frac{N}{2}$, and $d=|\frac{2J}{U}+\frac{N}{2}-k_0| \leqslant \frac{1}{2} $.
When $r=r_m$, previous results are recovered because $\frac{2J}{U}+\frac{N}{2} = -\frac{r_m}{2}$ is an integer and $\max[P(t)]=1$. For generic $r\neq r_m$, the lower bound of  $\max[P(t)]$ is written as $(1 + (\frac{\pi N U}{16 J})^2)^{-1/2} $. Thus, in the weakly interacting limit, Schr\"{o}dinger's cats well represent $\ket{\psi(t^{*\prime})}$. Away from $t=t^*$, we have numerically computed the overlaps between $\ket{\psi(t)}$ and the four states in Eq.\ \eqref{eq:states_and_times}, and such overlaps indeed reach their maxima near $t^*$ (Supplemental Materials). 

\begin{figure}
    \includegraphics[width=0.48\textwidth]{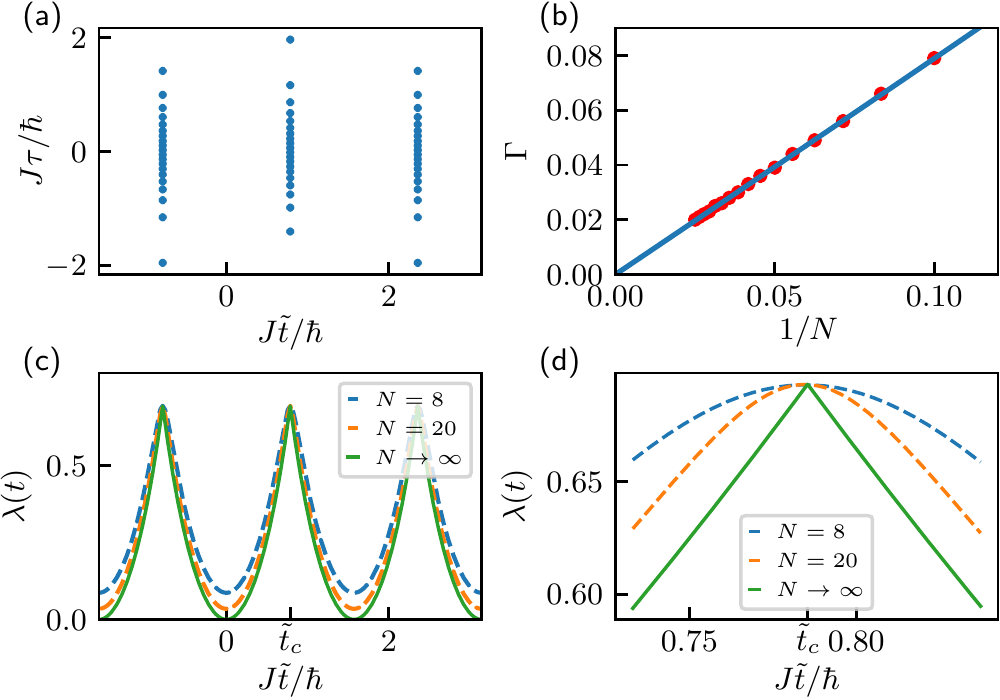}%
    \caption{\label{fig:four_figs} (color online)
    (a) Zeros of $G(z)$ near $t^*$ for $N=40$ particles ($\tilde t \equiv t-t^*$). (b) Distances between the real time axis and the nearest zeros around $t^*$ as a function of $1/N$. The blue line is the analytical result from Eq.\ \eqref{eq:bound_Gamma} and the red dots are numerical results. (c) The rate function $\lambda(t)$. (d) $\lambda(t)$ near 
    $\tilde t_c$. $UN^2/J=0.01$ have been used.}
\end{figure}

\paragraph{DQPT in the large $N$ limit. }  As explained before, in a short time scale of a few $T$s, the dynamics near $t^*$ is well captured by Eq.\ (\ref{eq:operator_substitution}) with the substitution $\tilde t=t-t^*$. Thus, the zeros of $G(z)$ in the complex plane can be obtained analytically near $t^*$. 
For instance, when $r=4m$,
\begin{equation}\label{eq:G(t)_for_r=4m}
G(z)=\frac{1}{\sqrt{2}}((\cos \frac{J\tilde z}{\hbar})^N-i(i\sin \frac{J\tilde z}{\hbar} )^N),
\end{equation}
where $\tilde{z}= \tilde{t}+i\tau$. As shown in Fig.\ \ref{fig:four_figs}(a), the real parts of these zeros are given by $\Re\tilde z=(\frac{\pi}{4}+\frac{m}{2}\pi)\frac{\hbar}{J}, m\in\mathbb{Z}$, i.e., these zeros are aligned in vertical lines in the complex plane. When $N$ is odd, some zeros reside on the real axis (Supplemental Materials). However, for a generic finite $N$, all zeros are away from the real axis.  With increasing $N$, zeros become denser and meanwhile gradually approach the real axis. In particular, the distance between the real axis and the nearest zero is bounded by 
\begin{align}\label{eq:bound_Gamma}
 \Gamma= \frac{1}{2}\arccosh{\frac{1}{|\cos\frac{\pi}{2N}|}}.
\end{align}
In the large $N$ limit, $\Gamma\approx\frac{\pi}{4N}$.  Such scaling behavior is verified by numerical calculations, as shown in Fig.\ \ref{fig:four_figs}(b). When $N\rightarrow \infty$, straight lines formed by continuous zeros intersect with the real axis and lead to a vanishing $G(z)$ in the real axis. Correspondingly, 
the rate function $\lambda(t)$ 
becomes nonanalytic, signifying DQPTs. 
As shown in Fig.\ \ref{fig:four_figs}(c,d), 
near the transition point, $\lambda(t)=\ln 2 - 2\frac{J}{\hbar}|\tilde t-\tilde t_c|$ when $N\rightarrow \infty$. 
 Comparing  DQPT points and the times given in Eq.\ \eqref{eq:states_and_times}, we conclude that pair condensates, $\ket{P_{\pm}}$, reside at DQPT points and characterize the DQPT between two different types of Schr\"{o}dinger's cats, $\ket{C_{\pm}}$.  This can also been seen from Fig.\ \ref{fig:zeros_and_gn}(c) and (f). Zeros of $G(z)$ near $t^*$ are aligned in a vertical line, which directly correspond to maximized $g_2$.

\paragraph{Effects of perturbations.}
Whereas essentially all parameters in Eq.\ \eqref{eq:H} can be fine tuned, it is useful to consider effects of perturbations. Here, we consider two types of important perturbations.
    (a) With increasing $U$, Eq.\ \eqref{eq:energy_first_order} includes high order terms $n^{s>2}$ . 
    (b) An energy mismatch $\Delta(n_1-n_2)$ breaks the inverse symmetry.

As for (a), the lowest order correction to the energy comes from a cubic term, $E_n=An+B n^2+Cn^3$, where  $Cn^3=-n^3 U^2 /(8 J)$ is given by the second order perturbation. Thus, the wave function is written as
\begin{align}
   \ket{\Psi(t)} = \sum_{n=0}^nc_n   e^{-\frac{i}{\hbar} (An+Bn^2- \frac{U^2}{8 J}n^3)t},\label{Pc}
\end{align}
where 
$c_n = (\frac{2}{\pi N})^{\frac{1}{4}}e^{-\frac{1}{N}(n-\frac{N}{2}-\frac{U}{16J}N^2)^2}$. 
If  $\frac{U^2}{8 J}n^3t^*\ll 1$ is satisfied, then the extra phase introduced by the cubic term is negligible within the time scale that is relevant to the emergent Schr\"{o}dinger's cat and DQPTs. All our previous results remain unchanged. Since $c_n$ is a Gaussian with a width $\sqrt{N}$,  which provides a natural cutoff of $n$ in the sum in Eq.\ (\ref{Pc}), we replace $n$ in the above inequality by $\sqrt{N}$ and obtain 
$UN^{\frac{3}{2}} \ll J$. 
The same discussions can be directly applied to higher order terms  $n^{s>3}$ in the energy.  Thus, when 
$UN^2\ll J$ is satisfied, 
all these corrections are negligible. 

Considering (b), 
our calculation (Supplemental Materials) shows that a finite $\Delta$ suppresses 
$g_N$ by a factor,
\begin{align}
    \frac{g_N}{g_{N}^0}=1-\left(\frac{\Delta^2N}{2J^2}+\frac{U\Delta N(N-1)}{16J^2}
    \right),
\end{align} 
where $g_N^0=N!/2$ is the $N$-body correlation function of a Schr\"{o}dinger's cat. Thus, when 
\begin{align}
    8\Delta^2N+\Delta UN(N-1)\ll 16J^2 \label{stability}
\end{align}
is satisfied, all characteristic features of a Schr\"{o}dinger's cat retain. 

It is interesting to compare Eq.\ (\ref{stability}) to the criterion for a stable Schr\"{o}dinger's cat at equilibrium. Whereas in the ideal situation, $\Delta=0$, a Schr\"{o}dinger's cat becomes the ground state when $U<0$, a finite $\Delta$ does not favor the superposition of $|N,0\rangle$ and $|0,N\rangle$, as a large $N$ amplifies the energy penalty. Meanwhile, the effective tunneling between $|N,0\rangle$ and $|0,N\rangle$ is exponentially small, as it requires $N$ steps of single-particle tunneling $J$ to couple these two states. Therefore, to access a Schr\"{o}dinger's cat as the ground state, it is required to have $\Delta N\ll Je^{-N}$, i.e., an exponentially small $\Delta$ with increasing $N$. This is the main obstacle to create a big cat state at the ground state when $N$ is large. Here, the constraint for the ground state does not apply to the Schr\"{o}dinger's cat generated in non-equilibrium quantum dynamics. Instead, Eq.\ (\ref{stability}) shows that, with increasing $N$, $\Delta$ only needs to be suppressed as a power law. In this sense, such dynamically generated Schr\"{o}dinger's cats are more stable than their counterparts at equilibrium. Thus, our results suggest a new route to access Schr\"{o}dinger's cats that can be potentially
used in precision measurements.

\paragraph{Experimental realizations.} Whereas our results apply to generic atomic interferometers, here, we comment on possible sceneries that are directly related to current experiments. A pair of optical tweezers has recently been used to create an atomic Hong-Ou-Mandel interferometer \cite{Regal2014Science}. Each single tweezer corresponds to a quantum state in Eq.\ \eqref{eq:H}. In such optical tweezers, both interaction $U$ and tunneling $J$ can be tuned. It is also possible to trap multiple atoms in a single optical tweezer \cite{LI2008135, Priv_Comm_Regal}. We have used realistic experimental parameters to verify that optical tweezers are indeed promising experimental platforms to explore DQPTs and emergent entangled states (Supplemental Materials).  Beside optical tweezers, other systems ranging  from double-well 
optical lattices to mesoscopic traps \cite{Sebby2006, Shin2004, Schumm2005, Islam2015}, in which the total particle number can be controlled precisely, are also suitable for testing our theoretical results. 

In summary, we have studied DQPTs in interacting atomic interferometers and shown that the dynamically generated entangled states have deep connections with zeros of Loschmidt echo in the complex plane. 
We hope that our work will stimulate more interests of using interactions in atomic interferometers as a constructive means to explore DQPTs and to produce novel entangled quantum states.

\begin{acknowledgments}
    This work is supported by startup funds from Purdue University. C.\/ Lyu also 
    acknowledges the support of F.\/ N.\/ Andrews Fellowship from Purdue University.
\end{acknowledgments}

\onecolumngrid

\newpage
\vspace{0.4in}

\centerline{\bf\large Supplemental Material}

\vspace{0.2in}

In this 
supplemental
material, we present the results of eigenstates and energy spectrum of the Hamiltonian, odd particle numbers, overlaps between the wave function and the four entangled states discussed in the main text, effects of perturbations, and optical tweezers. 

\section{Eigenstates and energy spectrum of the Hamiltonian}
We consider the Hamiltonian
 $\hat{H} = -J(\hat{a}_1^\dagger \hat{a}_2 + \hat{a}_2^\dagger \hat{a}_1) 
    + \frac{\bar{U}}{2}(\hat{n}_1+\hat{n}_2)^2+\frac{U}{4}(\hat{n}_1-\hat{n}_2)^2 + \Delta (\hat{n}_1 - \hat{n}_2)$. 
When $U=\Delta=0$,
the eigenenergies $E_n^0$
and eigenstates $\ket{E^0_n}$ 
are written as
\begin{align}
    E_n^0 = 2J(n-\frac{N}{2}),\,\,\,\,\,
    \ket{E_n^0} = 
    \frac{1}{\sqrt{n!(N-n)!}}(\frac{\hat{a}_1^\dagger + \hat{a}_2^\dagger}{\sqrt 2})^{N-n}
    (\frac{\hat{a}_1^\dagger - \hat{a}_2^\dagger}{\sqrt 2})^n\ket{0}.
\end{align}
When $U,\Delta\ll J$, the first and second order corrections 
to the eigenenergies are written as 
\begin{align}
    E_n^1 &= \frac{U}{4}(2nN-2n^2+N),\\
    \label{eq:energy_spectrum_2nd_correction}
    E_n^2 &= \frac{U^2}{32J}(2n-N)(N-1+2Nn-2n^2)+\frac{\Delta ^2}{2J}(2n-N).
\end{align}
The eigenstates are written as 
\begin{align}\label{eq:eigen_states_full}
    \ket{E_n} =& \ket{E_n^0} -\frac{\Delta}{2J}\sqrt{(n+1)(N-n)}\ket{E_{n+1}^0}
    +\frac{\Delta}{2J}\sqrt{n(N-n+1)}\ket{E_{n-1}^0} + O(\Delta^3)\\
    &  - \frac{U}{4}\frac{\sqrt{(N-n)(N-n-1)(n+1)(n+2)}}{4J}\ket{E_{n+2}^0}
    +\frac{U}{4}\frac{\sqrt{(n-1)n(N-n+1)(N-n+2)}}{4J}\ket{E^0_{n-2}}.
\end{align}

\section{Negative $U$}
As discussed in the main text,  when $U>0$, $t^*=\frac{\pi\hbar}{U}$, and 
$r_m=4m$, 
$\ket{C_-}$, $\ket{P_-}$, $\ket{C_+}$,
and $\ket{P_+}$ show up in order starting from $t^*$. 
In contrast, $r_m=4m+2$, 
$\ket{C_+}$,
$\ket{P_+}$, $\ket{C_-}$, and $\ket{P_-}$ show up in order starting from $t^*$. 
    
Here we discuss $U<0$ and $t^*=\frac{\pi\hbar}{|U|}$.
\begin{enumerate}
    \item $r_m=4m$,
    $\ket{C_+}$,
    $\ket{P_+}$, $\ket{C_-}$, and $\ket{P_-}$ show up in order starting from $t^*$, and $G(z)=\frac{1}{\sqrt{2}}((\cos J\tilde z/\hbar)^N+i(i\sin J\tilde z/\hbar)^N)$. 
    \item  $r_m=4m+2$, 
    $\ket{C_-}$, 
    $\ket{P_-}$, $\ket{C_+}$, and $\ket{P_+}$ show up in order starting from $t^*$, and $G(z)=\frac{1}{\sqrt{2}}((\cos J\tilde z/\hbar)^N-i(i\sin J\tilde z/\hbar)^N)$. 
\end{enumerate}
If $r_m$ is not an even integer, 
Eq.\ \eqref{eq:Pt_summation_form} can be generalized to 
\begin{align}
    P(t)\approx \sqrt{\frac{1}{1+\frac{N^2}{4}(\frac{\pi}{2}-\frac{|U|}{2\hbar}t)^2}}\times\sum_k 
        \left|\exp(-\frac{1}{\frac{2}{N}+i\frac{U}{|U|}(\frac{\pi}{2}-\frac{|U|}{2\hbar}t)}(\frac{k\pi}{2}-\frac{\pi N}{4} - \frac{Jt}{\hbar})^2 )\right |^2.
\end{align}

\section{Results for odd number of particles}
\begin{figure*}
    \includegraphics[width=\textwidth]{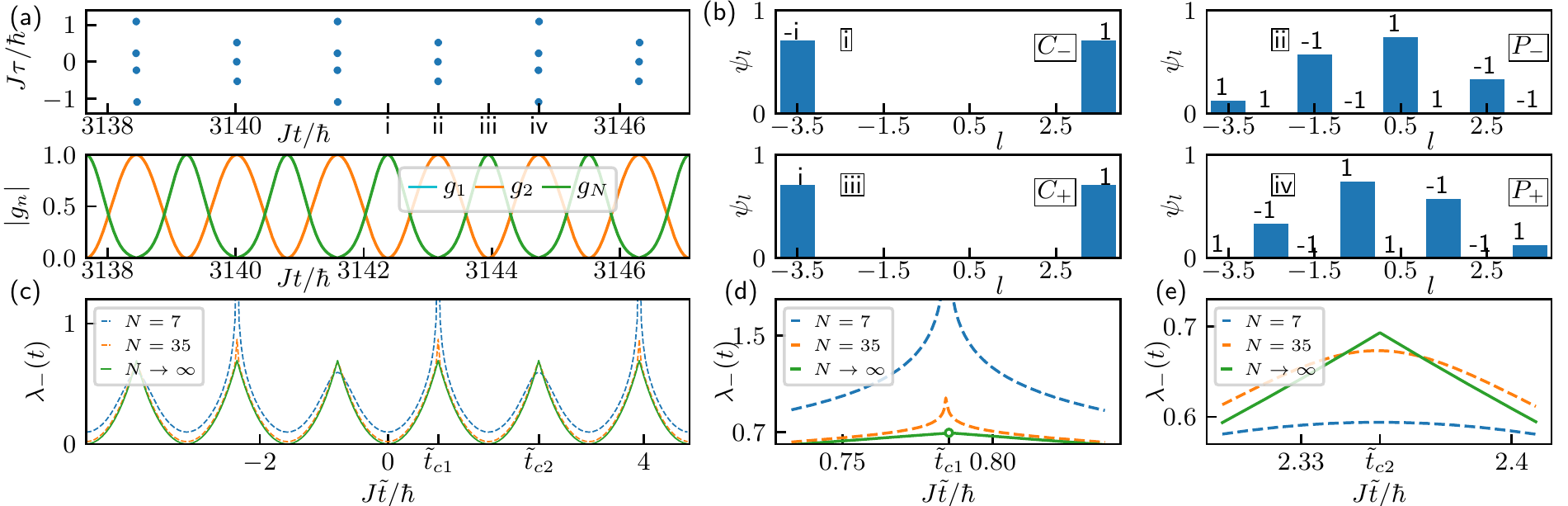}
    \caption{\label{fig:odd} (color online) (a) Zeros of $G(z)$ in the complex 
    plane of time for $7$ particles
    and the corresponding normalized correlation functions. (b) The wave functions at four
    times picked up from (a). (c) The rate function 
    $\lambda_{-}(t)$. 
    (d) and (e) show details of 
    $\lambda_{-}(t)$
    near $t_c$. The hollow dot in (d) represents the discontinuity of $\lambda_-(t)$
    at $\tilde t_{c1}$ where it approaches infinity. In all panels $U$ and $J$
    are fine tuned such that $\ket{\psi(t^*)} = \ket{C_-}$.}
\end{figure*}

The zeros of 
$G(z)=\frac{1}{\sqrt{2}}((\cos J\tilde z/\hbar)^N\pm i(i\sin J\tilde z/\hbar)^N)$
are written as
\begin{align}
    \Re\frac{J\tilde z}{\hbar} &= \frac{\pi}{4}+\frac{k}{2}\pi, k \in\mathbb{Z},\\
    \Im\frac{J\tilde z}{\hbar} &= \frac{1}{2}\arccosh\frac{1}{|\cos\pi(\frac{1+2k\mp1/2}{N}-\frac{1}{2})|}\cdot\sgn\sin \pi ( \frac{1+2k\mp1/2}{N}-\frac{1}{2})\label{eq:zero_im},\quad k = 1,2,..., N.
\end{align}
For a finite even $N$, zeros have finite imaginary parts. 
For a finite odd $N$, 
some zeros reside on the real time axis, as 
shown in Fig.\ \ref{fig:odd}.

In the large $N$ limit,
\begin{itemize}
    \item If $N$ is even, 
    $\lim_{N\rightarrow\infty}\lambda(t)=-2\ln[\max(|\cos J\tilde t/\hbar|, |\sin J\tilde t/\hbar|)]$, which has been
    analyzed in the main text.
    \item If $N$ is odd, $\lambda_{\pm}(t) = -\frac{1}{N}\ln(\frac{1}{2}|\cos^N J\tilde t/\hbar \pm \sin^N J\tilde t/\hbar|^2)$. The sign $\pm$ is determined 
    by the sign before $i$ in $G(t)$ and whether 
    $N=4p+1$ or $4p+3$, $p\in\mathbb{Z}$.
    $\lambda_{\pm}(t)$
    is nonanalytic at $\tilde t_{c} = \frac{\hbar}{J}(\frac{\pi}{4} + k\frac{\pi}{2}), k\in\mathbb{Z}$, when $N\rightarrow \infty$.
    Especially, $\lim_{N\rightarrow\infty}\lambda_-(t)=-2\ln[\max(|\cos J\tilde t/\hbar|, |\sin J\tilde t/\hbar|)]$
    except at $\tilde t_{c1}=\frac{\hbar}{J}(\frac{\pi}{4} + k\pi), k\in\mathbb{Z}$. As shown in 
    Fig. \ref{fig:odd}(d), $\lambda_-(t)$ diverges at $\tilde t_{c1}$ for any finite odd $N$. 
    Similar conclusions apply to $\lambda_{+}(t)$.

\end{itemize}

The emerged pair condensates near $t^*$ for odd $N$
are also different from those for even $N$, 
Using Eq.\ \eqref{eq:states_and_times}, 
for $N=2m+1, m\in\mathbb{Z}$, we obtain, 
\begin{align}
    \ket{P_-}&= \sum_{n=0}^N\frac{i^{N-n}-i^{n+1}}{p_n}\hat{a}^{\dagger n}_1\hat{a}^{\dagger N-n}_2\ket{0}
    = \sum_{n=0}^N\frac{ i^{n+1}( (-1)^{m+n} - 1  )   }{p_n}\hat{a}^{\dagger n}_1\hat{a}^{\dagger N-n}_2\ket{0},\\
    \ket{P_+}&= \sum_{n=0}^N\frac{i^{N-n}+i^{n+1}}{p_n}\hat{a}^{\dagger n}_1\hat{a}^{\dagger N-n}_2\ket{0}
    = \sum_{n=0}^N\frac{ i^{n+1}( (-1)^{m+n} + 1  )   }{p_n}\hat{a}^{\dagger n}_1\hat{a}^{\dagger N-n}_2\ket{0}.
\end{align}
Thus, some Fock states are suppressed by the 
factor $(-1)^{m+n} - 1$.  For instance when $N=7$, $\ket{P_-}$ only contains 
$\ket{0,7}, \ket{2,5}, \ket{4,3}, \ket{6,1}$. Apparently both one-body correction $g_1$ 
and $G(t) = \bra{7,0}\ket{P_-}$ vanishes.

\section{Overlaps between $|\psi(t)\rangle$ and $|C_{\pm}\rangle$, $|P_{\pm}\rangle$.}
\begin{figure*}
    \includegraphics[angle=0,width=0.9\textwidth]{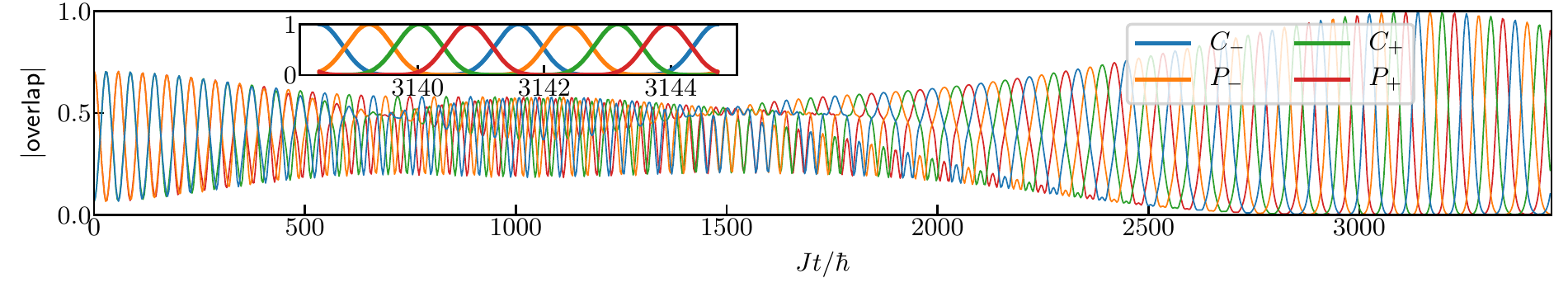}   
    \caption{\label{fig:overlap} (color online) The overlaps between the state $\ket{\psi_t}$ and four entangled states defined in the main text as a function of time. We have used $N=8$ bosons and $U/J=0.001$. }
\end{figure*}

Away from $t^*$, there is no simple analytical expression for the overlap between $|\psi(t)\rangle$ and the Schrodinger's cats or pair condensates. We thus evaluate such overlaps numerically, as shown in Fig.\ \ref{fig:overlap}. Near $t^*=\frac{\pi\hbar}{U}$, the four states defined in 
Eq.\ \eqref{eq:states_and_times}
show up alternatively. The overlaps reach maxima near $t^*$.

\section{Detailed analyses of perturbations}
When $\Delta=0$, 
the initial state $\ket{N,0}$ can be expanded by energy eigenstates and
the coefficients $c_n$
are 
\begin{align}
    |c_n|^2 = |\bra{E_n}\ket{N,0}|^2 = 
    |\frac{1}{2^{N/2}}\sqrt{\frac{N!}{n!(N-n)!}}(1-(N-2n)\frac{U(N-1)}{16J})|^2    \approx\sqrt\frac{2}{\pi N} e^{-\frac{2}{N}((n-\frac{N}{2})-\frac{UN(N-1)}{16J})^2}.
\end{align}
Assuming $E_n = Cn^3 + Bn^2+An$ and $B<0$, the overlap between $\ket{\psi_t}$
and the cat state $(\ket{N,0}+i\ket{0,N})/\sqrt 2$ is
\begin{align}
    \bra{C_+}\ket{\psi(t)} 
    &= \sum_{n=0}^N|c_n|^2e^{-i(\frac{\pi}{2}n^2+n\pi)}e^{-\frac{i}{\hbar}E_nt}
    = \sum_{m } \sqrt{\frac{2}{\pi N}}e^{-\frac{2}{N}m^2}
    \exp( -\frac{i C t}{\hbar} m^3-i(G_2-H_2t)m^2+i(G_1-\pi-H_1t)m).
\end{align}
where $m =n-\frac{N}{2}+\frac{UN^2}{16J}$ and 
\begin{align}
    H_2 &= \frac{|B|}{\hbar} - \frac{3CN^2U}{16\hbar J} -\frac{3CN}{2\hbar}, & G_2 &= \frac{\pi}{2},\\
    H_1 &= \frac{A}{\hbar}-\frac{|B| N^2 U}{8 \hbar J}-\frac{|B| N}{\hbar}+\frac{3 C N^4 U^2}{256 \hbar J^2}+\frac{3 C N^3 U}{16 \hbar J}+\frac{3 C N^2}{4 \hbar}, & G_1 &= -\frac{\pi N^2 U}{16J} - \frac{\pi N}{2}.
\end{align}
What is required is that the phase contributed by the cubic term
is negligible when $t\sim \frac{\pi\hbar}{U}$. Since the width of the gaussian factor is $\sqrt N$, we require
\begin{align}\label{speq:requirement_for_J}
    |\frac{C t}{\hbar} m^3| = |\frac{U^2}{8J}\frac{\pi}{U}N^{3/2}| \ll 1
    \Rightarrow |\frac{UN^{3/2}}{J}|\ll 1,
\end{align}
where we have used the energy spectrum obtained from second order perturbation. 
The cubic term is then dropped and we employ Poisson
summation formula to obtain,  
\begin{align}
    \bra{C_+}\ket{\psi(t)} =\sum_k \sqrt{\frac{1}{ 1+i\frac{N}{2}(G_2-H_2t) }}
    \exp(-\frac{1}{\frac{2}{N}+i(G_2-H_2t)}(\frac{(2k-1)\pi+ G_1}{H_1} - t)^2\frac{H_1^2}{4}).
\end{align}
Similarly, we obtain
\begin{align}
    \bra{C_-}\ket{\psi(t)} 
    =\sum_k \sqrt{\frac{1}{ 1+i\frac{N}{2}(G_2-H_2t) }}
    \exp(-\frac{1}{\frac{2}{N}+i(G_2-H_2t)}(\frac{(2k)\pi+ G_1}{H_1} - t)^2\frac{H_1^2}{4}).
\end{align}
When Eq.\ \eqref{speq:requirement_for_J} is satisfied, $H_2\approx \frac{|B|}{\hbar}
\approx \frac{U}{2\hbar}$, $H_1\approx \frac{A}{\hbar}\approx\frac{2J}{\hbar}$,
and $G_1\approx -\frac{\pi N}{2}$. We define the probability of finding a cat state as 
$P(t) = \max(|\bra{{C}_+}\ket{\psi(t)}|^2, |\bra{{C}_-}\ket{\psi(t)}|^2)$.
Near $t=\frac{G_2}{H_2}$, $P(t)$ can be written as 
\begin{align}
    P(t) &\approx \sqrt{\frac{1}{1+\frac{N^2}{4}(\frac{\pi}{2}-\frac{U}{2\hbar}t)^2}}\sum_k 
    \left|\exp(-\frac{1}{\frac{2}{N}+i(\frac{\pi}{2}-\frac{U}{2\hbar}t)}(\frac{k\pi\hbar}{2J}-\frac{\pi N\hbar}{4J} - t)^2 \frac{J^2}{\hbar^2})\right |^2.
\end{align}
$P(t)$ consists of multiple 
gaussian functions whose peaks are located at $t = \frac{k\pi\hbar}{2J}-\frac{\pi N\hbar}{4J}, k\in\mathbb{Z}$, 
and their separation is $\frac{\pi\hbar}{2J}$. There is also a factor 
$(1+\frac{N^2}{4}(\frac{\pi}{2}-\frac{U}{2\hbar}t)^2)^{-1/2}$,
which suppresses peak heights. 
If the parameters are fine tuned such that
an integer $k_0$ satisfies 
$\frac{\pi}{2}-\frac{U}{2\hbar}(\frac{k_0\pi\hbar}{2J}-\frac{\pi N\hbar}{4J}) = 0$, then
$P(t)=1$ at $t=\frac{k_0\pi\hbar}{2J}-\frac{\pi N\hbar}{4J}$. We thus obtain a perfect cat state.
Without fine tuning the parameters, we consider $t=\frac{\pi\hbar}{U}$ that lies in the 
middle of two peaks. The two peaks get a suppression of $(1 + (\frac{\pi N U}{16 J})^2)^{-1/2} $.
Again, because of Eq.\ \eqref{speq:requirement_for_J}, this factor is negligible when 
$N$ is large.

If the energy mismatch $\Delta$ is finite, 
we separate the eigenstates into two parts according to
their spatial parity,
\begin{align}
    \ket{E_n} = \alpha_n\ket{E_n}_s + \beta_n\ket{E_n}_a,   \quad  \hat P\ket{E_m} = \alpha_n(-1)^n\ket{E_n}_s + \beta_n(-1)^{n+1}\ket{E_n}_a.
\end{align}
The time evolution of the wave function is written as $\ket{N,0}\rightarrow \ket{\psi_t} = \sum_{n=0}^N c_n\alpha_n e^{-iE_nt}\ket{E_n}_s + c_n \beta_n e^{-iE_nt}\ket{E_n}_a$. 
From Eq.\ \eqref{eq:energy_spectrum_2nd_correction}, we see that, 
up to the second order of 
$\Delta$, the quadratic term in $E_n$ remains unchanged. Thus, when $t^*=\frac{\pi\hbar}{U}$, 
   $ e^{-iE_nt^*} = \frac{1+i(-1)^n}{\sqrt 2}$ is satisfied, and we obtain
\begin{align}
    \ket{\psi({t^*})} = \sum_{n=0}^N \alpha_n c_n \frac{1+i(-1)^n}{\sqrt 2}\ket{E_n}_s +
    \beta_n c_n \frac{1+i(-1)^{n}}{\sqrt 2}\ket{E_n}_a 
    =\ket{cat} + \ket{err}, 
\end{align}
where  $\ket{err}=\sum_{n=0}^Ni\sqrt 2(-1)^nc_n\beta_n\ket{E_n}_a$ is the correction to the cat state at $t^*$, and
\begin{align}
    g_N &= \bra{\psi_{t^*}}\hat{a}_1^{\dagger N}\hat{a}_2^N\ket{\psi_{t^*}} = g_N^0 + 
    \frac{N!}{\sqrt 2}\bra{0,N}\ket{err} + i\frac{N!}{\sqrt 2}\bra{err}\ket{N,0}
    + \bra{err}\hat{a}_1^{\dagger N} \hat{a}_2^N \ket{err}.
\end{align}
Using Eq.\ \eqref{eq:eigen_states_full}, we 
obtain 
\begin{align}
    \beta_n\ket{E_n}_a = -\frac{\Delta}{2J}\sqrt{(n+1)(N-n)}\ket{E_{n+1}^0}+\frac{\Delta}{2J}\sqrt{n(N-n+1)}\ket{E_{n-1}^0} + O(\Delta^3).
\end{align}
Up to the first order of $U$ and $\Delta$,
\begin{align}
    \ket{err}^{(1)} = \frac{\Delta}{2J}
    \sum_{n=0}^Ni\sqrt 2 \frac{(-1)^{n+1}}{2^{N/2}}\sqrt{\frac{N!}{n!(N-n)!}}(N-2n)\ket{E_n^0}.
\end{align}
It is straightforward to verify that $\bra{0,N}\ket{err}^{(1)}$, $\bra{err}^{(1)}\ket{N,0}$ and $\hat{a}_1^{\dagger N}\hat{a}_2^N\ket{err}^{(1)}$ vanish.

Up to the second order of $U$ and $\Delta$,
\begin{align}
    \ket{err}^{(2)} = \frac{\Delta}{2J}(\frac{\Delta}{2J}+\frac{U(N-1)}{16J}
    )\sum_{n=0}^Ni\sqrt 2 \frac{(-1)^{n+1}}{2^{N/2}}\sqrt{\frac{N!}{n!(N-n)!}}(N-2n)^2\ket{E_n^0},
\end{align}
\begin{align}
    \bra{0,N}\ket{err}^{(2)} 
    =-i\sqrt 2 \frac{\Delta}{2J}(\frac{\Delta}{2J}+\frac{U(N-1)}{16J}
    )\frac{N!}{(N-1)!}, \quad 
    \bra{err}^{(2)}\ket{N,0} = 0.
\end{align}
Therefore, 
\begin{align}\label{eq:g_N_D_U}
    g_N = \bra{\psi({t^*})}\hat{a}_1^{\dagger N}\hat{a}_2^N\ket{\psi({t^*})} = i\frac{N!}{2} 
    -i(\frac{\Delta}{2J}(\frac{\Delta}{2J}+\frac{U(N-1)}{16J}
    ))\frac{N!^2}{(N-1)!}
    = g_N^0(1-2N(\frac{\Delta}{2J}(\frac{\Delta}{2J}+\frac{U(N-1)}{16J}
    ))).
\end{align}

\section{Correlation functions and zeros of $G(z)$ in optical tweezers}
\begin{figure}
    \includegraphics[width=0.9\textwidth]{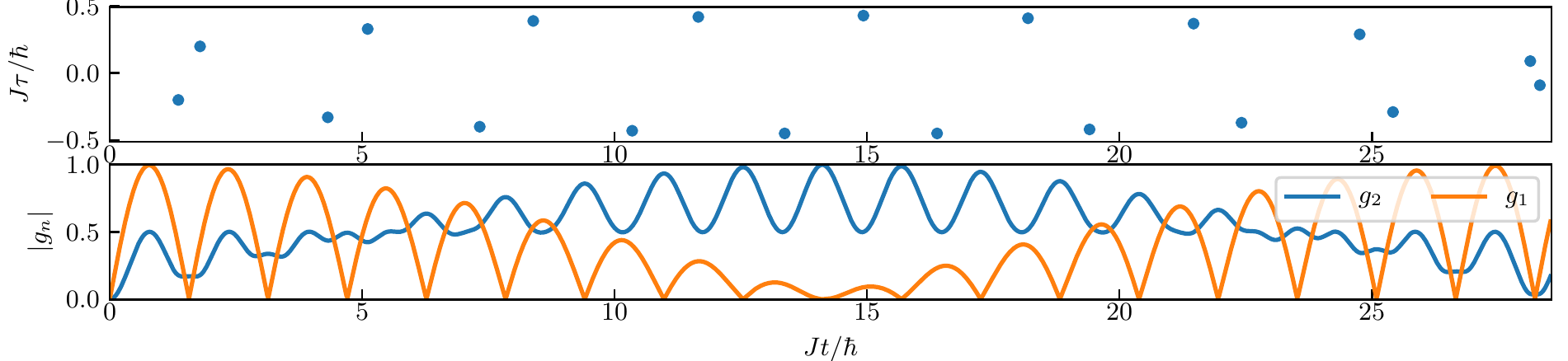}
    \caption{\label{fig:N2} Zeros of $G(z)$ in the complex plane and normalized correlation
    functions for 2 particles in optical tweezers. $U/J=0.22$.}
\end{figure}

Two coupled optical tweezers have been used to create an atomic Hong-Ou-Mandel interferometer \cite{RegalPRX, Regal2014Science}. Starting from an initial state, 
$\ket{2,0}$, i.e., two bosons occupy the same optical tweezer, the time evolution of the correlation functions can be calculated analytically,
\begin{align}
    g_1 &= -\frac{2U}{\sqrt{16J^2+U^2}} \alpha\beta\sin^2\frac{\sqrt{16J^2+U^2}t}{2\hbar}
    +i2\alpha\beta\sin\frac{\sqrt{16J^2+U^2}t}{2\hbar}\cos\frac{Ut}{2\hbar},\\
    g_2 &=  \frac{\alpha^4+\beta^4-1}{2} +\alpha^2\beta^2\cos\frac{\sqrt{16J^2+U^2}t}{\hbar}
    + i(\sin\frac{Ut}{2\hbar}\cos\frac{\sqrt{16J^2+U^2}t}{2\hbar}-\frac{U}{\sqrt{16J^2+U^2}}\cos\frac{Ut}{2\hbar}\sin\frac{\sqrt{16J^2+U^2}t}{2\hbar}),
\end{align}
where 
$
    \alpha = \frac{1}{\sqrt 2}\sqrt{1-\frac{U}{\sqrt{16J^2+U^2}}},
    \beta = \frac{1}{\sqrt{2}}\sqrt{1+\frac{U}{\sqrt{16J^2+U^2}}}.
$
If the parameters are fine
tuned such that $\frac{\sqrt{16J^2+U^2}}{U} = 2k, k\in\mathbb{Z}$, 
at $t^*=\frac{\pi\hbar}{U}$, we obtain, 
$
    g_1 = 0,g_2 = i(-1)^k
$,
and a small cat state $\frac{\ket{2,0}+i(-1)^k\ket{0,2}}{\sqrt{2}} $.
Using realistic experimental parameters in Ref.\ \cite{Regal2014Science}, the correlation functions and the zeros of $G(z)$ are
shown in Fig.\ \ref{fig:N2}.
When $U\ll J$, 
$\frac{\sqrt{16J^2+U^2}}{U} = 2k$ corresponds to 
$r=r_m$ in the main text. Without fine tuning experimental parameters, there are corrections to the small cat state at $t^*$, similar to the results discussed in the main text. It is worth mentioning that, starting from $|1,1\rangle$, the current experiment has shown that a small cat state can be produced in a Hong-Ou-Mandel interferometer. However, this is only true when interactions are ignored. We have verified that, in the presence of interactions, $|1,1\rangle$ cannot produce a small cat. Instead, $|2,0\rangle$ should be used, as shown by the previous discussions. 

\begin{figure}
    \includegraphics[width=0.9\textwidth]{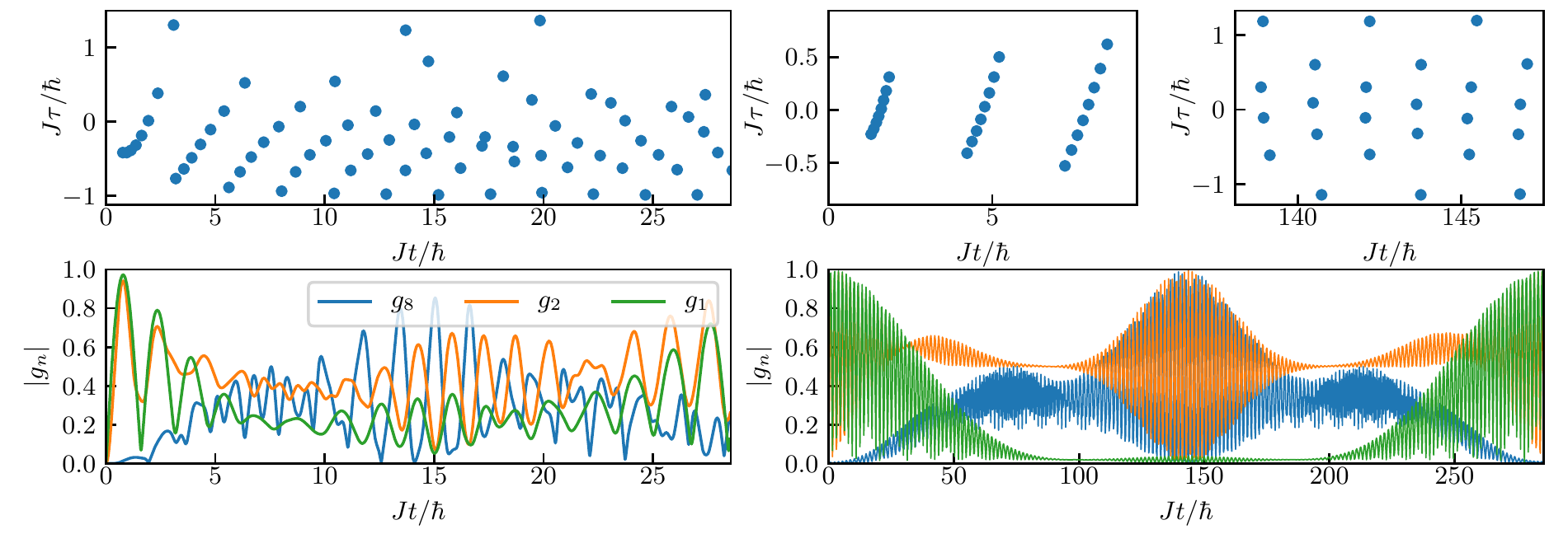}
    \caption{\label{fig:N8} Zeros of $G(z)$ in the complex plane and normalized correlation
    functions for 8 particles in optical tweezers. Left: $U/J=0.22$. Right: $U/J=0.022$.}
\end{figure}

It is possible that optical tweezers could trap multiple particles. For $8$ particles, $UN^2\ll J$ is no longer satisfied. Nevertheless, qualitative results remain unchanged. 
As shown in Fig.\ \ref{fig:N8}, $g_8$ is maximized near $t^*$ while other correlation functions are suppressed. With $U/J$ decreased down to $0.022$, all results in the main text are recovered. 

\end{document}